\documentclass[prc,aps,twocolumn,floats,floatfix,showpacs,superscriptaddress,nofootinbib]{revtex4-2}
\usepackage{amsmath}
\usepackage{amsfonts}
\usepackage{graphicx}
\usepackage{color}
\usepackage{comment}
\begin{document}
\title{Spreading widths of giant monopole resonance in the lead region:
Random matrix approach}
\author{N.N. Arsenyev}
\affiliation{Bogoliubov Laboratory of Theoretical Physics, Joint Institute for Nuclear Research, 141980 Dubna, Moscow Region, Russia}
\author{A.P. Severyukhin}
\affiliation{Bogoliubov Laboratory of Theoretical Physics, Joint Institute for Nuclear Research, 141980 Dubna, Moscow Region, Russia}
\affiliation{Dubna State University, 141982 Dubna, Moscow Region, Russia}
\author{R.G. Nazmitdinov}
\affiliation{Bogoliubov Laboratory of Theoretical Physics, Joint Institute for Nuclear Research, 141980 Dubna, Moscow Region, Russia}
\affiliation{Dubna State University, 141982 Dubna, Moscow Region, Russia}
\affiliation{rashid@theor.jinr.ru}
\date{\today}

\begin{abstract}
The microscopic calculation of the decay width of giant monopole resonance (GMR)  anticipates
 the mixing of one-phonon states with configurations of increasing complexity.
 To this aim we develop the effective approach for description of monopole excited states that
 are obtained in the quasiparticle random phase approximation (QRPA), with regard of the coupling
 between one- and two-phonon states. Based on the QRPA one-phonon states, we generate the coupling
 and two-phonon states by means of the Gaussian orthogonal ensemble (GOE) distribution. Within our
 approach the spreading width of the GMRs in $^{204,206,208}$Pb are described by means of a
 random matrix approach on two energy scales. It is demonstrated that the main contribution into
 the decay of the GMR is determined by a small number of two-phonon states strongly coupled to
 low-energy surface vibrations. While a vast majority of the coupling matrix elements (that are
 small in value and following the GOE distribution) are responsible for the fine structure of
 the GMR spreading width. A remarkable agreement between the results of the full microscopic
 calculations (based on QRPA phonons coupled by means of the microscopic coupling matrix elements
 with calculated two-phonon states) with those of the developed approach confirms the vitality of the proposed ideas.
\end{abstract}


\date{\today}

\maketitle

\section{Introduction}
Response of a finite quantum system to external excitations is one of the oldest
but still among most important subjects in quantum many-body theory.
Evidently, the increase of the excitation energy results in rise of
level density of the excited states of such a system.
Consequently, decay properties of the single-particle states and collective excitations of
a system under consideration may be described statistically above the particle emission
threshold, with a high degree of disorder (see in context of nuclear physics, e.g.,
Refs.\cite{brod,zel}).

Among various phenomena, related to this concept, the decay of giant
nuclear resonances (GRs) remains to be a topical subject in nuclear
structure theory during a few decades \cite{bertsch,wamb,S,bal,kam}.
GR states can be excited, for instance, by the nucleon-nucleus scattering,
by stripping of a nucleon from the projectile in the collision of two nuclei,
or by electromagnetic radiation. Contrary to the above statistical concept,
GRs  involve many nucleons in a coherent motion  and are characterized by
definite quantum numbers (spin, parity, isospin), rather than a
chaotic dynamics of uncorrelated particles.

Yet the analysis of nuclear collective properties of GRs requires as well the consideration of
their coupling with a stochastic background of compound states
(see, e.g., Refs.\cite{Lac,chav,rap,sav} and references therein).
According  to a general wisdom, the wave function of a particular GR is
rather spreaded over the eigenstates of the nuclear Hamiltonian, carrying the same
quantum numbers. In other words, in microscopic approaches, for a particular reaction
GR serves as a doorway state that is coupled to a set of background states via real
coupling matrix elements (see, e.g., Refs. \cite{BM,sol})
\footnote{Note that in deformed nuclei there are a few states due to a deformation splitting,
            which is beyond of our interest in the present paper.}.
As a result, such a state manifests itself as a broad maximum in the strength function
$S(E)=\sum_{k}|\langle k|F|0\rangle|^2\delta(E-E_k)$\,.
Here the matrix element $\langle k|F|0\rangle$ of the operator $F$, acting on the initial state
$|0\rangle$, creates the eigenstates of the nuclear Hamiltonian.
It is generally accepted that the decay evolution of the doorway states
over the hierarchy of more complex configurations to compound
states determines the spreading width $\Gamma^\downarrow$.
Together, with the Landau damping $(\Gamma_L)$ and
the escape width $(\Gamma^\uparrow)$, the spreading width forms the decay width
$\Gamma=\Gamma_L+\Gamma^\downarrow+\Gamma^\uparrow$ of a GR.
We recall that the Landau damping describes the fragmentation of one-particle
one-hole $(1p-1h)$ excitations, while the escape width corresponds to direct particle emission into the continuum.

With recent development of  semiconductor detectors
and computer facilities there is a desire to understand, at least,
the basic principles of decay mechanisms of various GRs, their common and distinctive properties.
The general idea on GR decay properties as a consequence of the coupling of high-lying modes
with the lowest collective vibrational modes \cite{bertsch} requires further
development in light of discussion on the role of order and disorder in nuclei \cite{hans,gom}.
We recall, however, that the analysis of spreading widths,
associated with the cascade of couplings and their fragmentations
due to these couplings (cf. Refs.\cite{jap1,lac,lacE,heiss}),
is a real challenge for nuclear structure theory.
In fact, even modern computer facilities are  unable to trace the decay of the doorway state
over the hierarchy due to the tremendous numerical obstacles. Nowadays, most successful attempts
in this direction are restricted by the consideration of the microscopic coupling between
one-phonon and two-particle-two-hole $(2p-2h )$ or two-phonon  configurations
(see, e.g.,  discussion in Refs.\cite{wamb,kam,ham,shev2,jap,tse2,nik1}).

In this paper we suggest the alternative approach, based on ideas of the
Random Matrix Theory (RMT) \cite{brod,met},
which enables us to count effectively the problem of the hierarchy at the description of
spreading widths. To provide a detailed overview of our approach
we consider only spherical or near-spherical nuclei around $^{208}$Pb, and focus our
attention on the decay width $\Gamma\approx\Gamma_{L}+\Gamma^{\downarrow}$ of the
GMRs. The escape width $(\Gamma^{\uparrow}$) is neglected in our approach,
since its contribution is negligible for heavy nuclei. It is noteworthy that a wide interest
to the decay properties of GMRs stems from the intention to extrapolate from these properties
the incompressibility of uniform nuclear matter (see for a review Refs.\cite{blai,garg,bul}).
More importantly for our discussion, that in the description of GMRs there is a need for
inevitable accounting  of the microscopic coupling between one-phonon and
$2p-2h$ configurations for correct interpretation of experimental data
(see details for a chain of nuclei in Ref.\cite{Li2023}).
Furthermore, it was stated that more complex configurations
may further improve the agreement with the experimental data.
We shall demonstrate that GMR spreading widths can be successfully simulated
by means of the RMT approach, based on the microscopic QRPA calculations.

\section{Theoretical Framework}

It appears that the main mechanism responsible for spreading widths differs for different GRs.
In particular, it was shown in Ref.\cite{shev2} that the coupling with the low-lying surface vibrations
provides quite satisfactory description of the width of the isoscalar quadrupole GR (ISQGR).
It seems that the Landau damping yields the major contribution to
the gross structure  of the isovector dipole GR (IVDGR)
\cite{it1,it2,it3}. However, the incorporation of ideas, borrowed from the RMT, providing the
effective counting of the two-phonon configurations, contributed additionally to
redistribution of the isovector dipole strength distribution \cite{sev4,sev5}.

Successful description of the IVDGRs within the RMT approach in the lead region
suggests to describe the GMR in the same vein.
To demonstrate the validity of our approach we shall compare the results of:
i) the microscopic calculations, based on the coupling between one-phonon and
two-phonon configurations, so called phonon-phonon coupling (PPC);
ii) the random matrix approach based on the one-phonon approximation;
iii) available experimental data for $^{204,206,208}$Pb nuclei.
To this aim we employ the modern development of
the quasiparticle-phonon model, where the single-particle
spectrum and the residual interaction
are determined making use of the Skyrme interaction
without any further adjustments~\cite{cme}.

Hereafter, we use the
parameter set of SLy4 \cite{22,22a}, which is adjusted to reproduce
the nuclear matter properties, as well as nuclei charge radii,
binding energies of doubly magic nuclei.
The pairing correlations are generated by a  zero-range volume force.
The pairing constant is taken as -280 MeV fm$^3$ \cite{sev4}.
In order to limit the pairing
single-particle space, we have used the smooth cutoff at 10~MeV
above the Fermi energies \cite{21}.
Below, for a self-contained discussion of our approach developed in Sec. \ref{newRMT},
 the main features of the PPC approach and the doorway model based on the RMT ideas
will be overviewed briefly.

\subsection{The PPC model}
\label{PPC}

By means of the finite rank separable approximation \cite{20,21}
for the residual interaction
we perform the QRPA calculations in very large two-quasiparticle spaces.
The cutoff of the
discretized continuous part of the single-particle spectra is taken
at the energy of 100 MeV. This is sufficient to exhaust practically
all the sum rules \cite{gg,nik1}. The QRPA solutions are treated as quasi-bosons with
quantum numbers $\lambda^\pi$. Among these solutions there are one-phonon
states $\omega_{\lambda i}$ corresponding to collective GRs and
pure two-quasiparticle states.

To construct wave functions of the excited $0^{+}$ states up to 20 MeV
we take into account all two-phonon terms that are built from the phonons of different
multipolarities $\lambda^{\pi}=0^{+},1^{-},2^{+},3^{-},4^{+}$,
coupled to $0^{+}$ state. In other words, we build two-phonon configurations that consist of
the phonon compositions $[\lambda_i^{\pi}\otimes \lambda_j^{\pi}]_{\lambda^\pi=0^+}$
(see details in Refs.\cite{rap,nik1,yad}). Following the basic ideas of the quasiparticle-phonon
model \cite{S}, the Hamiltonian is then diagonalized in a space spanned
by states composed of one and two phonons coupled by means of the
microscopic coupling matrix elements (see details in Refs.\cite{cme,sap12}).
The diagonalization results in eigenstates $|0^{+}_{\nu}\rangle$ with corresponding energies
$\omega_{\nu}$.

The basic steps of the calculations of the strength distribution of the GMR, $b(E0,E)$,
in the PPC approach for $^{208}$Pb are discussed in \cite{yad}.
In brief, we define the strength distribution as
\begin{equation}
b(E0,E)=\sum_{\nu}|\langle 0^{+}_{\nu}|{\hat M}_{\lambda=0^+}|0^+\rangle|^2\rho(E-E_\nu)\,,
\end{equation}
where $|\langle 0^{+}_\nu|{\hat M}_{\lambda=0^+}|0^+\rangle|^2$ is the transition probability
from the ground state $|0^+\rangle$ to
the excited state $|0^{+}_\nu\rangle$. The transition operator of the GMR is defined as
\begin{equation}
\label{mop}
{\hat M}_{\lambda=0^+}=\sum_{i=1}^A r_i^2\,.
\end{equation}
The strength distribution is described with the aid of the Lorentzian function
\begin{equation}
\rho(\omega-\omega_\nu)=\frac{1}{2\pi}\frac{\Delta}{(\omega-\omega_\nu)^2+\Delta^2/4}
\end{equation}
with $\Delta=1$ MeV.

The QRPA analysis provides the location of the GMR in  $^{204,206,208}$Pb
in the energy region $E_{x}=8-20$~MeV.
In particular, in nucleus $^{208}$Pb there is  one, strongly dominating peak in the strength
distribution at 14.6 MeV (see Fig.~\ref{fig:OneScale}).
The PPC yields a detectable redistribution of the GMR strength in
comparison with the RPA results. It results in
the 1~MeV downward shift of the main peak. Our analysis shows that the major
contribution to the strength distribution is brought about by the
coupling between the $[0^+]_{RPA}$ and $[3^-\otimes3^-]_{RPA}$
components \cite{yad}.

\subsection{The RMT approach}
\label{rmt}

Let us recapitulate the basic steps of the statistical description of the
GMR fragmentation based on ideas from the RMT \cite{sev4,sev5}.
In our approach the one-phonon states are generated by means of the QRPA calculations,
while the coupling matrix elements between the one-phonon and two-phonon states
are replaced by random matrix elements of the GOE-type. Namely, we consider a doorway Hamiltonian
\begin{equation}
H_{\lambda^\pi}=H_d+H_b+V\,,
\end{equation}
where the Hamiltonian
$H_d=\sum\limits_i^{N_d} \omega_i Q^+_i Q_i$
is characterised by
energies $\omega_i$ obtained from the microscopic calculations of the monopole
phonon states, and the $N_d$ one-phonon states constitute the doorway states.
The background states $N_b$ are two-phonon and possibly more complex states,
are eigenstates of the Hamiltonian
$H_b=\sum\limits_k^{N_b} \Omega_k a^+_k a_k$
with eigenstates $|b;\Omega_k\rangle$ and corresponding energies $\Omega_k$. The
number of background states is much larger than the number
of doorway states, $N_b{\gg}N_d$.

\begin{figure}[htb]
\includegraphics[width=0.4\textwidth]{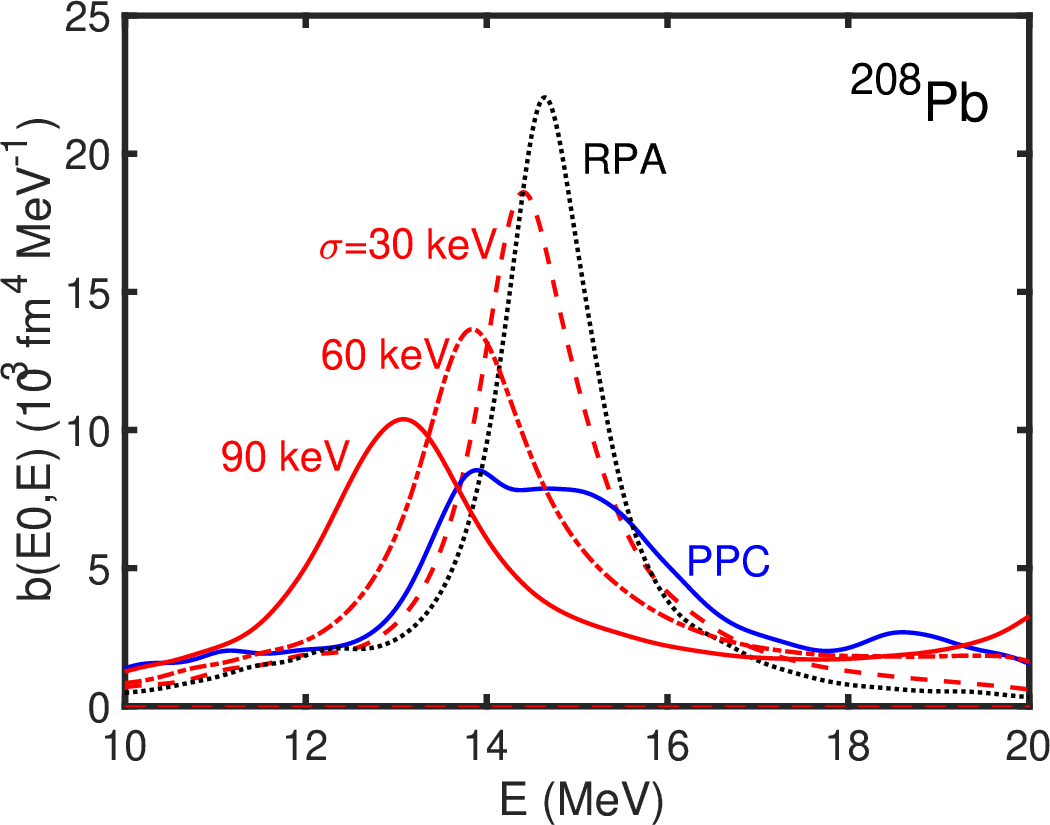}
\caption{E0 strength function, $b(E0,E)$, for $^{208}$Pb as a function of energy.
The dotted (black) line connects the RPA results, and the solid (blue) line connects the PPC results.
The random matrix approach on one energy scale is applied for
three values of the coupling strengths, $\sigma$=30 keV (red dashed line),
60 keV (red dot-dashed line), and 90 keV (red solid line).
}
\label{fig:OneScale}
\end{figure}

We recall that the Hamiltonian $H_{\lambda^\pi}$ represents a set of good
quantum numbers, $\lambda^{\pi}$, and the QRPA phonons as well as all background states
fulfill these quantum numbers. We assume
no coupling between different doorway states or between different background states,
$\langle d|V|d' \rangle=0$ and $\langle b|V|b'\rangle =0$,
but all coupling takes place between the doorway states and
the complex background states,
$\langle d |V| b \rangle=V_{db}\Rightarrow V_{d_i,b_k}= \langle d;\omega_i|V |b;\Omega_k \rangle$.
As discussed above, the microscopic coupling (PPC)
matrix elements are replaced by a random interaction where the matrix elements
$V_{d_i,b_k}$, are Gaussian distributed random numbers,
\begin{equation}
P(V_{d_i,b_k})=\frac{1}{\sigma\sqrt{2\pi}} \exp\left({\frac{-V_{d_i,b_k}^2}{2\sigma^2}}\right)\,,
\label{rand}
\end{equation}
with the width or strength
$\sigma=\sqrt{\langle V_{d_i,b_k}^2\rangle}$
and fulfilling
$V_{d_i,b_k}=V_{b_k,d_i}$
The one-phonon states are thus considered as doorway states to
the fragmentation of $E0$-strength on background states.

Thus, our aim is to describe  microscopically the
one-phonon GMRs states, and attempt a random matrix inspired
treatment of the coupling to complex surrounding states, here viewed as two-phonon
states. The quality of the random treatment can then be studied by comparing results
with the microscopic PPC model predictions.
The use of the random matrix distribution yields the
backshifting of the main peak. In fact, with increasing the coupling strength
$\sigma=$ 30, 60 and 90 keV the peak of the strength distribution of monopole excitations
is gradually pushed down in strength in case of $^{208}$Pb
(see Fig.~\ref{fig:OneScale}). At the same time the peak is pushed down
to lower excitation energies. Note, in this case there is only an average
strength that does not produce any preferences in the coupling between
one- and two-phonon states of different one-phonon nature.
There seems to be no way to come close to the PPC result in the RMT model for any coupling strength.
The reason for this is that the density of two-phonon states increases with excitation energy,
as $\rho_{2-ph} \propto E^3$ \cite{sev5}. With a random Gaussian distributed coupling
between the  one-phonon and the two-phonon states, the number of two-phonon states
with high energies are much larger than the number of the matrix elements coupled to low two-phonon energies.
Consequently, the increase of the repulsion (the coupling strength $\sigma$) between the one-phonon and two-phonon states pushes
the main peak of the strength function down in energy, while the high-lying tail
appears as well. This picture deviates more and more from the PPC result as the coupling strength $\sigma$ is increased.

\begin{figure}[ht]
\includegraphics[width=0.4\textwidth]{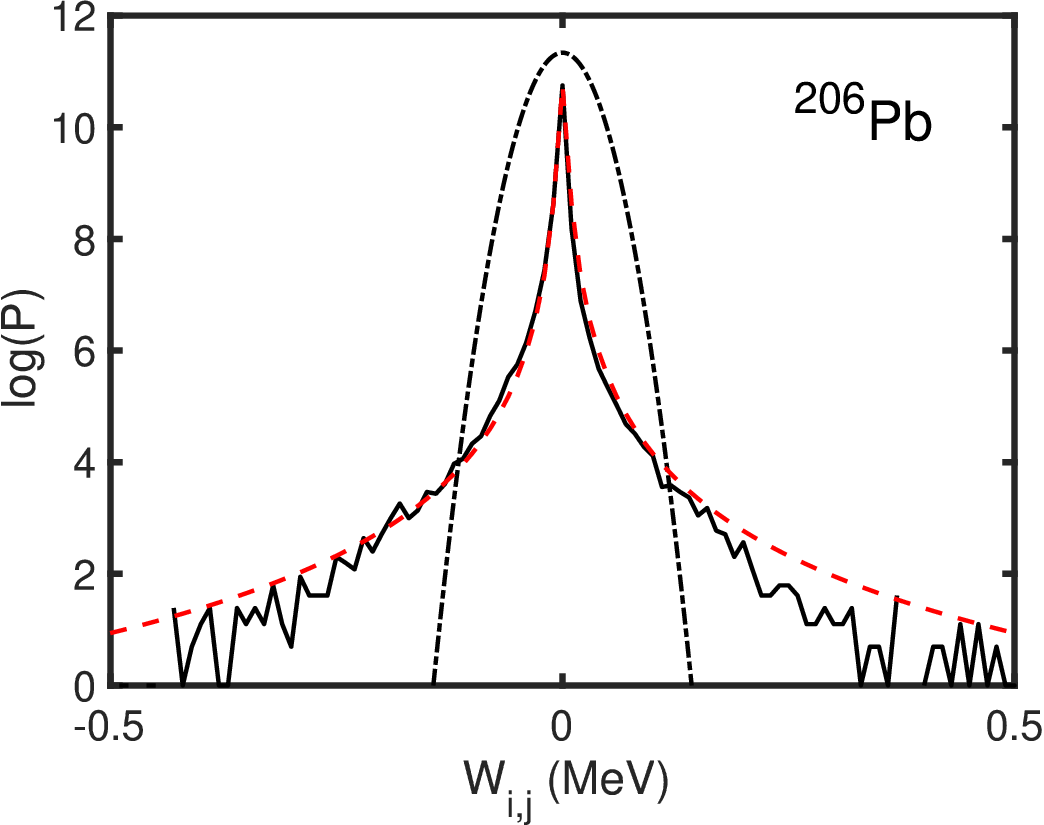}
\caption{Distribution of the coupling matrix elements between monopole one-phonon
and two-phonon QRPA states. The probabilities $P(W_{i,j})$ are shown on log scale
for $^{206}$Pb (black solid line). A fit to a Cauchy distribution
is shown by red dashed line. The Gaussian distribution of the matrix elements, used
for the weak coupling with $\sigma_2$=30 keV, is shown by a dot-dashed line.
}
\label{fig:Dist}
\end{figure}

\section{The RMT on different energy scales}
\label{newRMT}

In order to resolve the arising problem we turn to the successful attempts
to understand the fine structure of the ISQGR with the aid of the wavelet
analysis \cite{shev2}. In the heart of this analysis there is the idea
on presence of different energy scales, responsible for fluctuations of the
cross section of any resonance (see also analysis within simple models in Refs.\cite{jap1,heiss}).
The essence of the analysis \cite{shev2} guides us to suppose that it would be useful to study
the decomposition of the full model space on different subspaces,
responsible for different decay mechanisms. In practical terms it
means the separation of the coupling matrix elements on two classes:
i) the coupling matrix elements responsible for the strong
coupling between low-lying vibrations and the doorway states;
ii) the coupling matrix elements between the doorway states and
large background of incoherent states.

To illuminate this suggestion we consider the distribution of
microscopic coupling matrix elements $W_{i,j}$ (calculated in the PPC approach)
for $^{206}$Pb, taken as a typical example (see Fig.~\ref{fig:Dist}).
The distribution of all coupling matrix elements is well reproduced by a (truncated) Cauchy distribution.
We recall that stable distributions, such as the Cauchy distribution, have long tails and infinite variance.
However, considering truncated Cauchy distributions,
according to the central limit theorem, the resulting shape (the average of the sum)
is driving the Gaussian distribution. Indeed, the distribution of the bulk of coupling matrix
elements (basically, small in value) follow approximately the Gaussian distribution.
While large coupling matrix elements are distributed on irregular tails.
In the energy interval $8-20$ MeV the number of one-phonon  QRPA states  are 27, 28 and 15,
 while the number of two-phonon states coupled to $0^+$ are 2126, 2210 and 914 for
 $^{204}$Pb, $^{206}$Pb and $^{208}$Pb, respectively.
The rms-value of the many matrix elements corresponding to the Gaussian distribution
(the central part in Fig. \ref{fig:Dist}) is approximately
$\sigma_2\approx 30$ keV for $^{204,206,208}$Pb,
which we shall use in the random matrix approach (see below).

\begin{figure}[ht]
\hspace*{-0.5cm}
\includegraphics[width=0.65\textwidth]{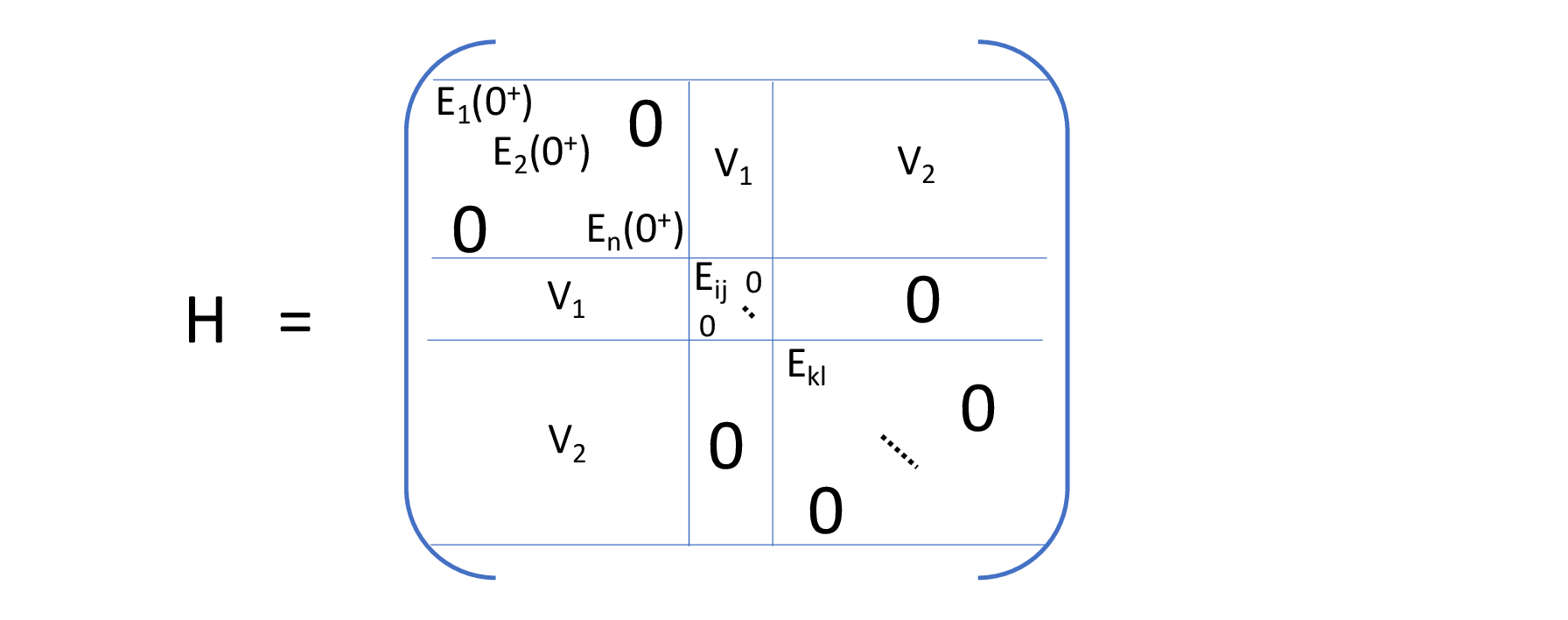}
\caption{
Schematic view of the random matrix Hamiltonian on two energy scales.
First submatrix describes the one-phonon QRPA states with energies $E_i(0^+)$.
Middle submatrix is associated with the two-phonon states with energies $E_{i,j}$ that
are strongly coupled to the one-phonon states with the random coupling $V_1$.
Third submatrix is associated with the two-phonon states with energies  $E_{k,l}$
that are weakly coupled to the one-phonon states with the random coupling $V_2$.
There is no coupling between one-phonon states, or between two-phonon states.
Only one-phonon and two-phonon states are coupled.}
\label{fig:Ham}
\end{figure}

Taking the above analysis into account, we model
the random matrix Hamiltonian in the following way:
 \begin{equation}
 \hat{H}=\hat{H}_0 + \hat{H}_1 + \hat{H}_2\,.
 \label{2scales}
 \end{equation}
Here the term $\hat{H}_0$ describes doorway states. These states are associated with
the $0^+$ one-phonon states calculated within the QRPA.
The two-phonon Hamiltonians are
\begin{equation}
\hat{H}_k=\hat{H}_{k0}+\hat{V}_k, \hspace{0.2cm} k=1,2\,,
\end{equation}
where $\hat{H}_{k0}$ describes two-phonon states that consist of the phonon compositions
as \mbox{$[\lambda_i^{\pi}\otimes \lambda_j^{\pi}]_{\lambda^\pi=0^+}$,}
with energies $E_{ij}=E(\lambda_i)+E(\lambda_j)$, built by means of one-phonon
states with $\lambda^\pi$=$0^+$, $1^-$, $2^+$, $3^-$ and $4^+$.
The  term  $\hat{H}_{10}(\hat{H}_{20})$ is coupled strongly (weakly) to
the one-phonon states by means of the random force $\hat{V}_{1(2)}$.
In this case the strong (weak) coupling matrix elements  are Gaussian random variables
with zero mean value and a second moment $\sigma_1^2(\sigma_2^2)$
between the doorway states and the background states. The two-phonon states that belong to the term
$\hat{H}_{10}$ have a low-level density, typically 1 state per MeV.
The states that belong to the term $\hat{H}_{20}$ correspond to the rest of
the two-phonon states. The matrix representation of the discussed Hamiltonian is displayed on  Fig.\ref{fig:Ham}.

At this point there are a few comments in order.
Thus, we have to determine the main principle how to select the strongest
coupling matrix elements for our random matrix approach.
To this aim we propose the procedure without performing the full PPC calculations,
where the matrix elements actually are calculated.
The QRPA provides the energy and the electric transition matrix element
of the one-phonon state, corresponding to a given multipolarity $\lambda^\pi$,
to the ground state $|0^+\rangle$, i.e.,
$B(E\lambda)\propto |\langle \omega_{\lambda^\pi}|{\hat E}(\lambda)|0^+\rangle|^2$.
Evidently, the most collective vibrations (phonons) of different multipolarity are
of main interest for us.
Therefore, the electric transition matrix element of each phonon state is subsequently transformed
to Weisskopf single-particle units, $B_{s.p.}(\lambda)_i$
by dividing the calculated transition strength by the Weisskopf estimate \cite{BM}
for transition from the ground state $0^+$ to the excited states $\lambda^\pi$.
On the other hand, keeping in mind the fact that two-phonon states are
 formed from the tensor product of identical multipole operators
counted at different energies, we consider the joint product of
their transition probabilities to the ground state
$M_{i,j}=B_{s.p.}(E\lambda)_i \cdot B_{s.p.}(E\lambda)_j$.
\begin{table}
\caption{Characteristics of ten
largest matrix elements
between a two-phonon state coupled to $\lambda^{\pi}=0^+$,
and the one-phonon RPA $0^+$ state at 14.6 MeV carrying the strongest
monopole transition strength to the ground state in $^{208}$Pb.
First column: the configuration of the two-phonon states is given in terms
of two coupled QRPA phonons.
Second column: the corresponding energies.
Third column: the corresponding PPC matrix elements.
Fourth column: the product of transition matrix elements, $M_{i,j}$,
in terms of single-particle Weisskopf units (see text).
}
\begin{ruledtabular}
\begin{tabular}{ccccccccccccccccc}
2-ph state & E$_{2ph}$ (MeV)  & $|W_{i,j}|$ (MeV) & $M_{i,j}$/100  \\
\noalign{\smallskip}\hline\noalign{\smallskip}
$[3^-_1\otimes 3^-_1]_{0^+}$    &  7.1 & 1.37 & 26.87\\
$[2^+_1\otimes 2^+_1]_{0^+}$    & 10.4 & 0.33 &  0.86\\
$[3^-_6\otimes 3^-_1]_{0^+}$    & 10.5 & 0.35 &  0.84\\
$[4^+_1\otimes 4^-_1]_{0^+}$    & 11.2 & 0.34 &  2.89\\
$[3^-_9\otimes 3^-_1]_{0^+}$    & 11.3 & 0.29 &  0.93\\
$[4^+_5\otimes 4^+_1]_{0^+}$    & 14.2 & 0.38 &  1.63\\
$[4^+_6\otimes 4^+_1]_{0^+}$    & 14.8 & 0.28 &  0.96\\
$[3^-_{41}\otimes 3^-_1]_{0^+}$ & 16.1 & 0.58 &  0.01\\
$[2^+_{15}\otimes 2^+_1]_{0^+}$ & 18.1 & 0.68 &  2.00\\
$[3^-_{56}\otimes 3^-_1]_{0^+}$ & 18.2 & 0.34 &  0.91\\
\end{tabular}
\end{ruledtabular}
\end{table}

\begin{figure}[ht]
\hspace*{-0.5cm}
\includegraphics[width=0.4\textwidth]{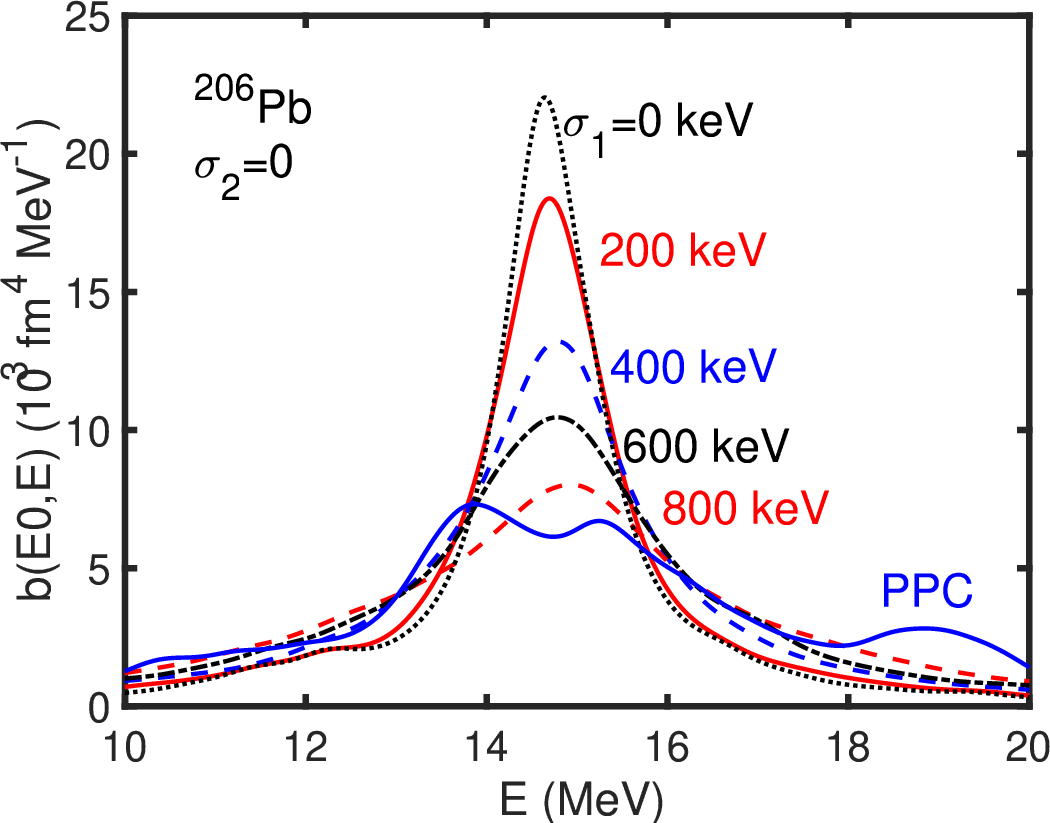}
\caption{The monopole transition strength $b(E0,E)$
versus the transition energy E in the case of $^{206}$Pb.
The results are shown for different values
of the strong coupling strength, $\sigma_1$, assuming the week coupling strength
$\sigma_2=0$. For a  comparison the result, obtained by means of the
PPC theory, is shown by blue solid line.}
\label{fig:Sigma1}
\end{figure}

In Table I the ten largest microscopic coupling matrix elements, obtained
by means of the PPC approach, are compared with the $M$-values for the
corresponding two-phonon compositions in the case of $^{208}$Pb.
It can be seen that the selection of the two-phonon states with the largest
$M$-values is in fair agreement with the largest values of $|W_{i,j}|$.
Out of the two-phonon states with the ten strongest matrix elements, nine are
found by this simplified rule.
The only exception is the two-phonon state with $E=16.1$~MeV,
which is built on the connection of the first $3^{-}_{1}$~vibration
with the pure two-quasiparticle $3^{-}_{41}$~state.
Additionally, two-phonon states
with two $0^+$ RPA phonons, coupled to $0^+$, appear to have very small
coupling matrix elements with the one-phonon $0^+$ RPA phonons, and are neglected in this process.
The reason why the two-phonon states with two  $0^+$ phonons give a small
coupling to the one-phonon  $0^+$ state is a geometrical factor, as described in Appendix B of Ref.\cite{cme}.

The ten energies, shown in Table I, are evenly distributed over the considered
energy interval with about one state per MeV, and the rms-strength of the
PPC calculated coupling matrix elements is $\sigma_1= 588$ keV.
Similar results are obtained for $^{204,206}$Pb with the rms values
of the ten strongest matrix elements $\sigma_1=560, 590$ keV, respectively.
Thus, given these results, we shall consider  the group of strongly coupled matrix elements
with the coupling strength $ \sigma_1\approx$600 keV and with the density $\rho_1$=1 MeV$^{-1}$.
\begin{figure}[ht]
\centering
\includegraphics[width=0.4\textwidth]{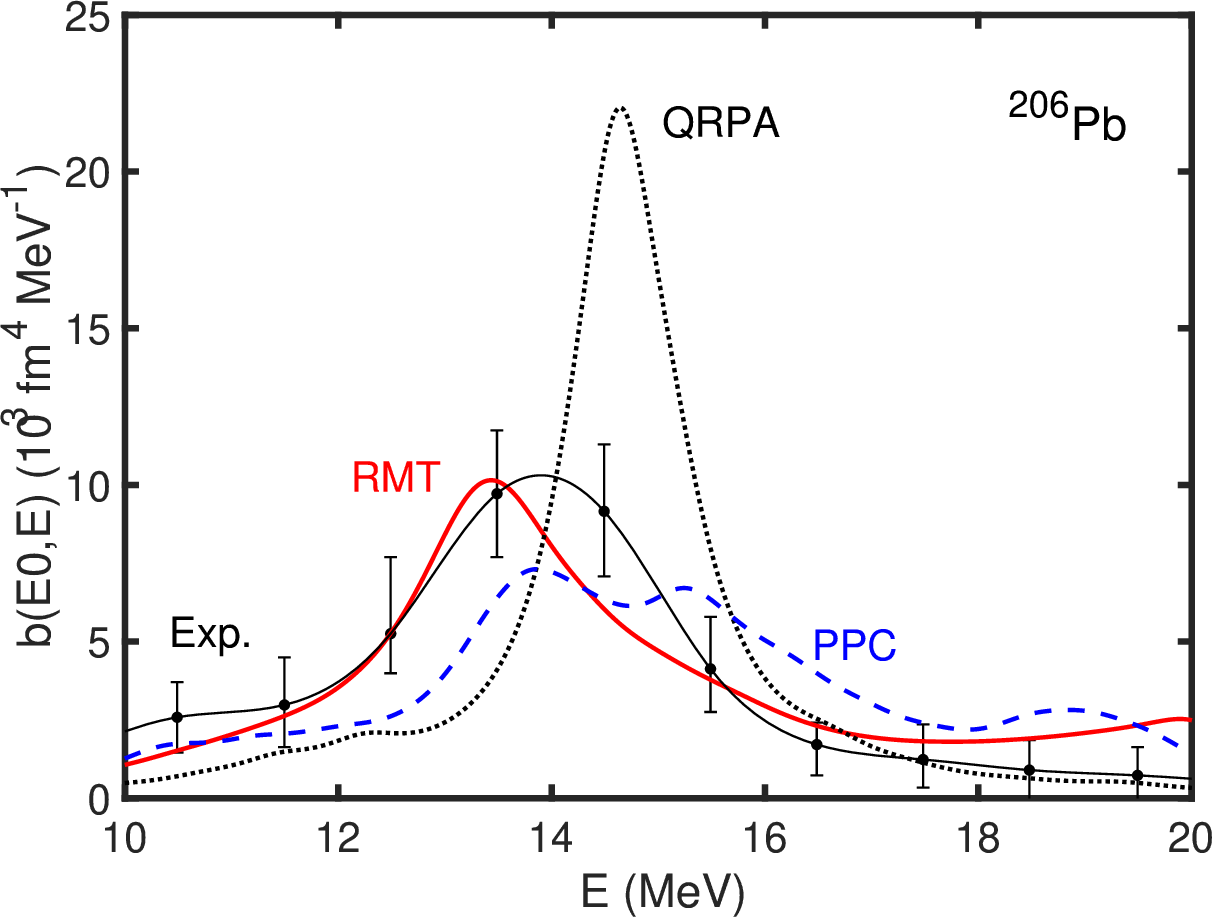}
\caption{The monopole transition strength $b(E0,E)$
versus the transition energy E in the case of $^{206}$Pb.
The results, obtained by means of: i) the two-scale RMT approach are connected by (red) solid
line; ii) the microscopic PPC calculations are  connected by (blue) dashed line;
iii)the QRPA approach are connected by (black) dotted line. For a comparison
the experimental data \cite{P13} shown by (black) squares with error bars, smoothly interpolated, are
connected by  (black) thin line.}
\label{fig:206Pb}
\end{figure}

Fig.\ref{fig:Sigma1} shows the strength distributions of monopole excitations for different values
of the random coupling matrix elements strength $\sigma_1$ between the subspace of two-phonon states
and the one-phonon GMR state.
These strength distributions are
obtained by ensemble averaging over 100 realizations. Each realization is calculated
with the aid of the coupling matrix elements
randomly generated by means of the Gaussian distribution (\ref{rand}).
We recall that  Eq.(\ref{rand}) approximates (according to the central limit theorem) the
resulting shape (the average of the sum)
random generations of the coupling matrix elements
for each considered value $\sigma_1$.
If the weak part of the interaction is neglected in the RMT model ($\sigma_2$=0),
the strength function is gradually broadened as the strong interaction ($\sigma_1$)
is increased (see Fig.~\ref{fig:Sigma1}). Switching on  the strong as well as the
week interactions, with the chosen values $\sigma_1$=600 keV and $\sigma_2$=30 keV,
the RMT results are in a quite good agreement with those of the PPC (see Fig.~\ref{fig:206Pb}).
It is notable that the strength distribution of the GMR, obtained in this case,
is rather close to the experimental distribution~\cite{P13}.

\begin{figure}[ht]
\centering
\includegraphics[width=0.4\textwidth]{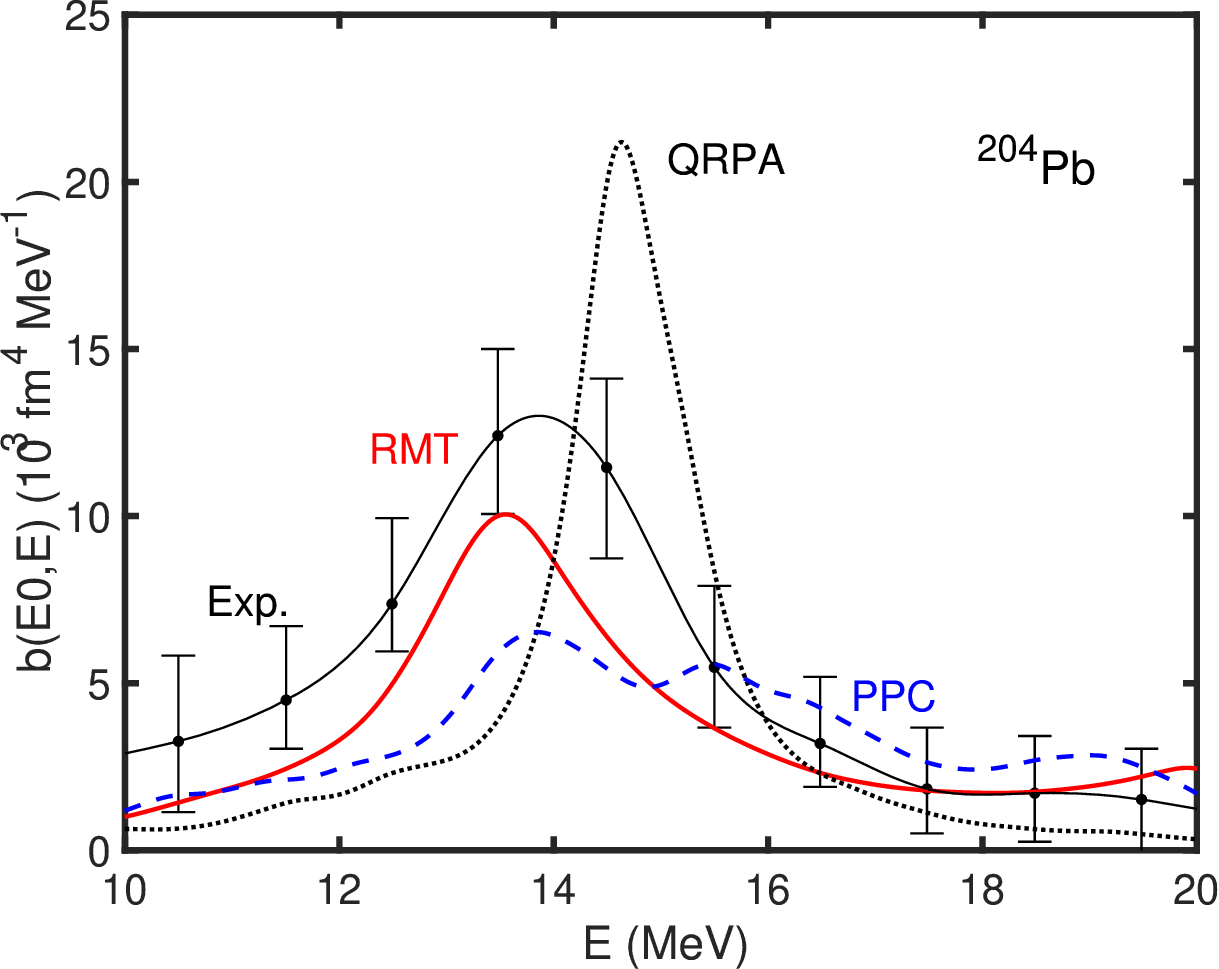}
\caption{Similar to Fig.~\ref{fig:206Pb} but for $^{204}$Pb.
}
\label{fig:204Pb}
\end{figure}

\begin{figure}[ht]
\centering
\includegraphics[width=0.4\textwidth]{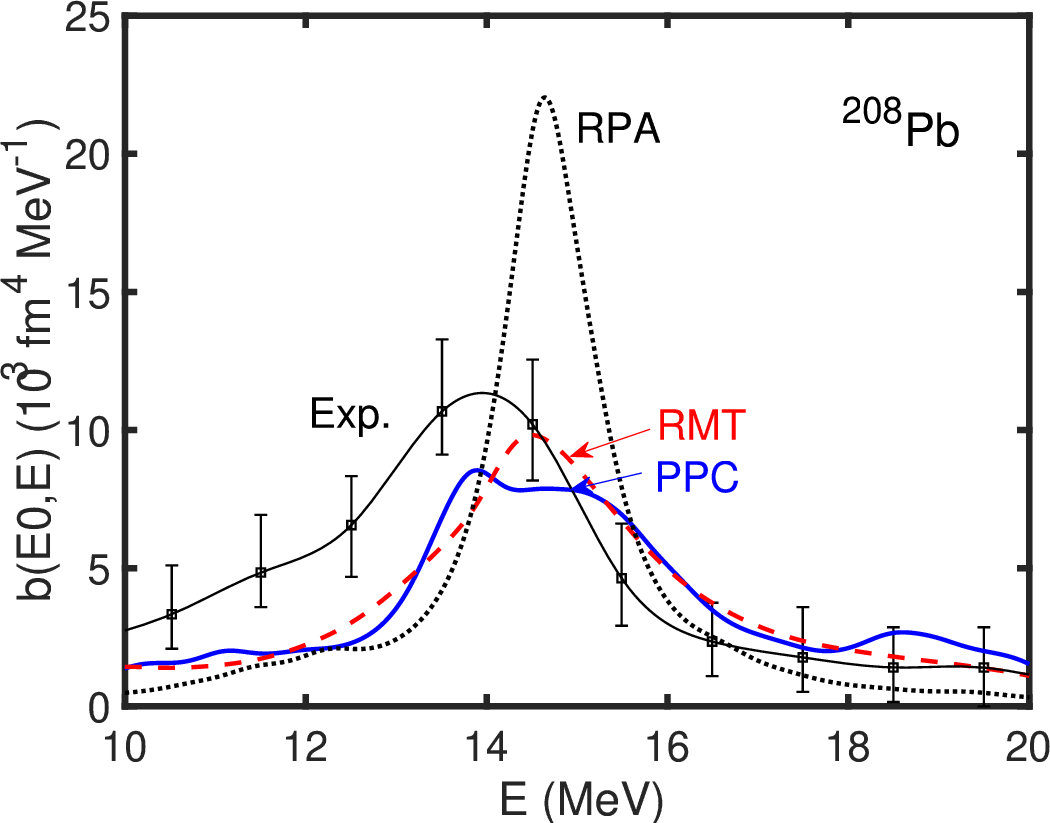}
\caption{Similar to Fig.~\ref{fig:206Pb} but for $^{208}$Pb.
The RMT results are connected by dashed (red) line, while the PPC results
are connected by solid (blue) line.}
\label{fig:208Pb}
\end{figure}

The experimental and calculated positions of the energy centroids, $E_c$ and
the width of the GMRs for the three considered lead isotopes are shown in Table II.
The values of $E_{c}$ and $\Gamma$
have been computed in the energy region $10.5-18.5$~MeV and defined
by means of the energy-weighted moments $m_k=\int_{}b(E0,E)\,{E}^k\,dE$:
i) $E_{\rm c}=m_1/m_0$; ii) $\Gamma=2.35\sqrt{m_2/{m_0}-\left(m_1/{m_0}\right)^2}$.
Applying the RMT approach, based on two energy scales, to  the
calculation of the monopole  strength
distribution for $^{204,208}$Pb ,
we obtain quite satisfactory agreement with the experimental data \cite{P13}.
On the other hand, the agreement between the results of the microscopic PPC  and the RMT
calculations  of the monopole strength distribution
(see Figs.~\ref{fig:206Pb}, \ref{fig:204Pb}, \ref{fig:208Pb}) is quite remarkable.

\begin{table}
\caption{Lorentzian fits of measured \cite{P13} and calculated in
the RMT, PPC approaches the values of centroid energy $E_c$
and width $\Gamma$ of E0 strength
function $b(E0, E)$, shown in Figs. \ref{fig:206Pb}, \ref{fig:204Pb}, \ref{fig:208Pb}.
The values of $E_{c}$ and $\Gamma$ have been computed in the energy region $10.5-18.5$~MeV.
}
\begin{ruledtabular}
\begin{tabular}{ccccccccccccccccc}
&\multicolumn{3}{c}{$E_{c}$ (MeV)}  &\multicolumn{3}{c}{$\Gamma$ (MeV)}  \\
&Experiment &\multicolumn{2}{c}{Theory}  &Experiment  &\multicolumn{2}{c}{Theory}\\
&                       &PPC  &RMT  &            &PPC &RMT\\
\noalign{\smallskip}\hline\noalign{\smallskip}
$^{204}$Pb&13.8$\pm$0.1 &14.7 &14.4 &3.3$\pm$0.2 &4.5 &4.3 \\
$^{206}$Pb&13.8$\pm$0.1 &14.6 &14.5 &2.8$\pm$0.2 &4.3 &4.2 \\
$^{208}$Pb&13.7$\pm$0.1 &14.6 &14.7 &3.3$\pm$0.2 &4.1 &4.0 \\
\end{tabular}
\end{ruledtabular}
\end{table}

\section{Summary}
In this paper we have suggested the effective approach for description of
the GMR spreading width. It is shown that this description can be successfully
fulfilled by means of the microscopic calculations of the QRPA states alone,
that are mixed by means of the coupling matrix elements with two-phonon states,
generated with the aid of the GOE distribution.

Within the framework of our approach the two-phonon model space is decomposed
on two subspaces that are differently coupled to the QRPA states.
We have demonstrated that two energy scales, corresponding to large and small
coupling strengths of two-phonon states to one-phonon states,  provide a better description
naturally, accounting for fine structure effects. On the larger energy scale
the gross structure and structure effects of the GMRs are defined; that includes
the random coupling to surface vibrations of a few strongest coupling matrix elements.
The weaker coupling on a smaller
energy scale is also important, and is particularly responsible for the fine
structure of the strength function in the energy region around the GMR.
Similar studies were performed in Ref.~\cite{shev2} for the ISQGRs,
where the characteristic scales have been discussed, motivated from a wavelet analysis
of the measured strength functions. In contrast to the latter case, we provided the
recipe of the selection of the most important coupling matrix elements that determine
the large scale of the strength (see Sec.\ref{newRMT}).

To illustrate the quality of our approach,
all numerical calculations have been done on the basis of the Skyrme forces SLy4.
Our major goal was to elucidate the efficiency of the proposed approach than rather to reproduce with
the high accuracy experimental data by selecting, for example, the specific Skyrme functional (e.g., Ref.\cite{Li2023}).
Consequently, our attempts were aimed to obtain a better understanding of the gross structure of the resonance
and, basically,  to reproduce the microscopic PPC result by modest means. The remarkable agreement between
the results of the PPC  and the RMT calculations  for the GMR strength distribution of $^{204,206,208}$Pb
confirms the vitality and validity of our approach.
Noteworthy is the fact that our approach can be readily extended with the proposition of the mixing with three-phonon states.
While it is a laborious task in microscopic calculations.

\begin{acknowledgments}
We are grateful to our friend Sven \AA berg for
the constant fruitful cooperation over the past years.
All results, presented in this paper, have been discussed in detail with
him. His criticisms led to significant improvements of the original
version of the manuscript, which began two years ago.
N.N. Arsenyev acknowledges the financial support from the
Russian Science Foundation (Grant No. RSF-21-12-00061).
\end{acknowledgments}

\end{document}